\documentstyle[epsf,aps]{revtex}
\begin{document}
\draft
\preprint{YITP-00-17,KUNS-1657}
\title{Thick Domain Walls Intersecting a Black Hole}
\author{ 
Yoshiyuki Morisawa$
{}^{1}$\footnote{E-mail:morisawa@yukawa.kyoto-u.ac.jp}, 
Ryo Yamazaki$
{}^{2}$\footnote{E-mail:yamazaki@tap.scphys.kyoto-u.ac.jp}, 
Daisuke Ida$
{}^{2}$\footnote{E-mail:ida@tap.scphys.kyoto-u.ac.jp}, 
Akihiro Ishibashi$
{}^{1}$\footnote{E-mail:akihiro@yukawa.kyoto-u.ac.jp}
 and 
Ken-ichi Nakao$
{}^{3}$\footnote{E-mail:knakao@sci.osaka-cu.ac.jp}
        } 
\vskip5mm
\address{${}^{1}$
Yukawa Institute for Theoretical Physics, Kyoto University,
Kyoto 606--8502, Japan, 
        }
\address{${}^{2}$
Department of Physics, Kyoto University,
Kyoto 606--8502, Japan, 
        }
\address{${}^{3}$
Department of Physics, Osaka City University,
Osaka 558--8585, Japan
         }

\date{\today}
\maketitle

\begin{abstract}
We discuss the gravitationally interacting system of a thick domain
wall and a black hole.
We numerically solve the scalar field equation in the Schwarzschild
spacetime and show that there exist scalar field configurations
representing thick domain walls intersecting the black hole.
\end{abstract}

\section{Introduction}

Topological defects arise during spontaneous symmetry breaking
associated with phase transitions, and cosmological evolution of them
is considered to have played an important role in
cosmology (see {\it e.g.}, \cite{VilenkinBook}).
Topological defects produced in the early universe might give us some
information on high energy phenomena which cannot be reached by
accelerator experiments, and those produced at the late time phase
transitions \cite{Hill:1989vm} also have attracted attention as a potential source of the cosmic structures.
To study the topological defects is therefore crucially important in
cosmology and elementary particle physics.

In general relativity, topological defects have interesting features.
Topological defects are extended and relativistic objects due to their
large tension.
In particular, the gravitational field produced by an infinitely thin
domain wall shows repulsive nature
\cite{Vilenkin:1983hy,Ipser:1984db}.
On the other hand, although there are many studies about the
properties of thick domain walls in the flat and de~Sitter spacetime
backgrounds \cite{VilenkinBook}, little is known about the existence
of thick wall configurations on inhomogeneous, strongly curved
background such as a black hole spacetime.
Thus it is intriguing to study the gravitational interaction between
two extended relativistic objects: a topological defect and a black
hole.

In most studies of defect--black hole system, topological defects have
been approximately treated as infinitely thin and non-gravitating
objects whose dynamics is governed by Nambu-Goto action.
Within this context, the scattering problem of a Nambu-Goto string by
a background black hole has been studied in detail
\cite{DeVilliers:1998nk,DeVilliers:1998xz,Page:1998ya,DeVilliers:1999nm,Page:1999fd}.
In the domain wall case, recently M.~Christensen, V.~P.~Frolov and
A.~L.~Larsen \cite{Christensen:1998hg,Frolov:1999td} considered
Nambu-Goto walls embedded in the Schwarzschild black hole spacetime
and found the static axisymmetric solutions.
They showed that there exist a family of infinitely thin walls 
which intersect the black hole event horizon.
However, little attention has been given to the system of a thick
defect interacting with a black hole though a defect as a
topologically stable configuration of a scalar field has a finite
thickness.

Now we shall consider the validity of thin-wall approximation in the
system of a topological defect and a black hole.
In such a system, there are two characteristic scales: the thickness $w$
of the defect and the black hole radius $R_g$.
In the case of the system of an astrophysical black hole with the mass
$\sim M_{\odot}$ and a defect formed during a GUT phase transition,
the thickness of the defect is much smaller than the black hole radius
and therefore thin-wall approximation would be valid.
However, it is not so hard to consider the situation that the
thickness of defects cannot be negligible as compared with the size of 
a small black hole.
Over the last few decades, many people have studied the formation
of small black holes called primordial black holes (PBHs) and their
cosmological implications.
For example, studying the contribution of
PBHs to cosmic rays enables one to place limits on the spectrum of
density fluctuations in the early universe
(see {\it e.g.} \cite{Cline:1998ft,Carr:YKIS99,Yokoyama:YKIS99}).
On the other hand, it has been discussed the possibility of thick
defects and their roles in cosmology, {\it e.g.} as a source of
large-scale structure in the universe \cite{Hill:1989vm}, or as a
candidate for some kind of dark matter \cite{Battye:1999eq}.
For example, it is thought that the typical mass of PBHs
which evaporate at the present epoch is $10^{15}\mbox{\rm g}$, so
$R_g\sim10^{-13}\mbox{\rm cm}$.
When one considers the topological defects formed during a phase
transition at $\lesssim 100\mbox{\rm MeV}$, such defects become
thicker than the size of PBHs and thin-wall approximation is no longer
valid.

In this paper, we investigate the gravitational interaction between a
domain wall and a black hole taking the thickness of the wall into
account.
We deal with scalar fields in the Schwarzschild black hole spacetime
with $\phi^4$ and sine-Gordon potentials, which have a discrete set of
degenerate minima.
We explicitly show that static axi-symmetric thick domain walls
intersecting the black hole do exist by numerical investigation.
We consider a non-gravitating domain wall for simplicity.
This test wall assumption might be valid when the symmetry braking
scale of the scalar field is much lower than Planck scale as will be
shown later by dimensional analysis.

This paper is organized as follows.
In section 2, we derive the basic equation and discuss the boundary
conditions which represent the situation we want to study.
In section 3, we show the numerical result.
We summarize our work in section 4.
We also discuss the validity of the assumption that the effects of
gravity of domain wall can be ignored near the black hole horizon.
Throughout this paper, we use units such that $c=\hbar=G=1$ unless otherwise 
stated.

\section{The basic equation and the boundary conditions}

We consider a static thick domain wall in a black hole spacetime.
The domain wall is constructed by a scalar field with self-interaction 
in a given curved spacetime.
In what follows, we neglect the self gravity of the scalar field, as
will be justified later.
As a background spacetime, we consider the Schwarzschild black hole
\begin{equation}
g=
-\left(1-\frac{2M}{R}\right){\rm d}t^2
+\left(1-\frac{2M}{R}\right)^{-1}{\rm d}R^2
+R^2({\rm d}\vartheta^2+\sin^2\vartheta {\rm d}\varphi^2).
\end{equation}
For our purpose, we find that it is more convenient to work in the
isotropic coordinates $\{t,r,\vartheta,\varphi\}$, where the new
radial coordinate $r$ is defined by
\begin{equation}
R=r\left(1+\frac{M}{2r}\right)^2.
\end{equation}
We mainly consider the region outside the event horizon in this paper,
which corresponds to $r>M/2$.
In this coordinate system, the metric has a spatially conformally flat
form
\begin{equation}
g=
-\left(\frac{2r-M}{2r+M}\right)^2{\rm d}t^2
+\left(1+\frac{M}{2r}\right)^4
\left[{\rm d}r^2+r^2({\rm d}\vartheta^2+\sin^2\vartheta {\rm d}\varphi^2)\right].
\end{equation}

Let us consider a real scalar field $\phi$ with a potential $V[\phi]$, 
of which Lagrangian is given by
\begin{equation}
{\cal L}=
-(-\det g)^{1/2}\left(\frac{1}{2}\nabla\phi\cdot\nabla\phi+V[\phi]\right).
\end{equation}
The equation of motion for $\phi$ is
\begin{equation}
\nabla^2\phi-\partial V/\partial \phi=0.
\label{eq:EOM}
\end{equation}
In this paper, we consider following two familiar types of potentials
which have a discrete set of degenerate minima; the $\phi^4$ potential
\begin{equation}
V_1[\phi]=\frac{\lambda}{4}(\phi^2-\eta^2)^2,
\end{equation}
 and
the sine-Gordon potential
\begin{equation}
V_2[\phi]=\lambda\eta^4[1+\cos(\phi/\eta)].
\end{equation}

Since the Schwarzschild spacetime is asymptotically flat, the
asymptotic boundary condition for Eq.~(\ref{eq:EOM}) would be given by
the solution in the flat spacetime.
The relevant solutions in the flat spacetime
$g=-{\rm d}t^2+{\rm d}x^2+{\rm d}y^2+{\rm d}z^2$ are the static and plane-symmetric
solutions
\begin{equation}
\phi_1(z)=\eta\tanh\sqrt{\lambda/2}\eta z
\label{eq:flat4}
\end{equation}
and
\begin{equation}
\phi_2(z)=\eta[4\arctan \exp(\sqrt{\lambda}\eta z)-\pi],
\label{eq:flatSG}
\end{equation}
for the potentials $V_1$ and $V_2$, respectively.
These solutions represent domain walls in the flat spacetime
characterized by the thickness of the wall
\begin{equation}
w = 1/\sqrt{\lambda}\eta.
\end{equation}
In the Schwarzschild background, the solution compatible with the
above asymptotic boundary condition would have a static and
axi-symmetric form $\phi=\phi(r,\vartheta)$.
Then, the explicit form of the equation of motion (\ref{eq:EOM})
becomes
\begin{equation}
\left(\frac{2r}{2r+M}\right)^4
\left[
\frac{\partial^2}{\partial r^2}
+\frac{8r}{(4r^2-M^2)}\frac{\partial}{\partial r}
+\frac{1}{r^2}\left(\frac{\partial^2}{\partial\vartheta^2}
+\cot\vartheta\frac{\partial}{\partial\vartheta}\right)
\right]\phi
=
\frac{\partial V}{\partial\phi}.\label{eq:explicit}
\end{equation}
This equation can be parameterized by a single dimensionless parameter
\begin{equation}
\epsilon=M/2w,
\end{equation}by introducing dimensionless variables
\begin{equation}
\rho=2r/M,~~~\Phi(\rho,\vartheta)  =  \phi(r,\vartheta)/\eta.
\end{equation}
In terms of these variables, Eq.~(\ref{eq:explicit}) becomes
\begin{equation}
\left(\frac{\rho}{\rho+1}\right)^4
\left[
\frac{\partial^2}{\partial \rho^2}
+\frac{2\rho}{(\rho^2-1)}\frac{\partial}{\partial \rho}
+\frac{1}{\rho^2}\left(\frac{\partial^2}{\partial\vartheta^2}
+\cot\vartheta\frac{\partial}{\partial\vartheta}\right)
\right]\Phi
=
\epsilon^2\frac{\partial U}{\partial \Phi},
\label{eq:basic}
\end{equation}
where the dimensionless potential
$U[\Phi]=V[\phi]/\lambda\eta^4$ is defined. 
$U$ has minima at $\Phi=\pm 1$ for $\phi^4$ potential and at
$\Phi=(2n+1)\pi$ ($n=0,\pm1,\pm2,\cdots$) for sine-Gordon potential.
The parameter $\epsilon$ is just a ratio of the horizon radius to the
wall thickness measured in the asymptotic region, namely if $\epsilon$
is smaller (larger) than unity, then the wall is said to be thick
(thin) as compared to the size of the black hole.

We shall confine ourselves to the case that the core
of wall is located at the equatorial plane $\{\vartheta=\pi/2\}$ of
the black hole.
The solutions without this assumption will be discussed in the
separated paper \cite{MYII2}.
Accordingly, we impose the Dirichlet boundary condition at the
equatorial plane
\begin{equation}
\Phi|_{\vartheta=\pi/2}=0.
\label{eq:BCequator}
\end{equation}
Now it is sufficient to consider the north region, namely 
the solution in the
south region can be obtained via 
$\Phi(\rho,\vartheta)=-\Phi(\rho,\pi-\vartheta)$,
$\{\pi/2 \le\vartheta\le\pi\}$
 of the spacetime.
The regularity of the scalar field at the symmetric axis is give by
the Neumann boundary condition
\begin{equation}
\frac{\partial\Phi}{\partial\vartheta}|_{\vartheta=0}=0.
\label{eq:BCaxis}
\end{equation}
On the other hand, the boundary condition at the event horizon
$\{\rho=1\}$ is given by the Neumann boundary condition
\begin{equation}
\frac{\partial\Phi}{\partial\rho}|_{\rho=1}=0.
\label{eq:BChorizon}
\end{equation}
As is shown in Appendix, the condition (\ref{eq:BChorizon}) is the
consequence of a natural requirement that the energy density observed
by a freely falling observer remains finite at the event horizon.
In practice, the region of numerical integration should be finite, so
that we need an asymptotic boundary condition at $\rho=\rho_{\rm max}$ 
for $\rho_{\rm max}\gg 1$.
Taking into account the flat background solutions (\ref{eq:flat4}) and
(\ref{eq:flatSG}), we impose the Dirichlet boundary condition
\begin{equation}
\Phi_1|_{\rho=\rho_{\rm max}} 
= \tanh (2^{-1/2}\epsilon\rho_{\rm max}\cos\vartheta)
\label{eq:BCasymptotic1}
\end{equation}
and
\begin{equation}
\Phi_2|_{\rho=\rho_{\rm max}} =4\arctan \exp(\epsilon\rho_{\rm max}\cos\vartheta)-\pi,
\label{eq:BCasymptotic2}
\end{equation}
for the $\phi^4$ and the sine-Gordon potentials, respectively.

In the next section, we numerically integrate the field equation
(\ref{eq:basic}) using relaxation method under these boundary
conditions at the equatorial plane (\ref{eq:BCequator}), at the
symmetric axis (\ref{eq:BCaxis}), at the event horizon
(\ref{eq:BChorizon}) and in the asymptotic region
(\ref{eq:BCasymptotic1}), (\ref{eq:BCasymptotic2}) for both the
$\phi^4$ and the sine-Gordon potentials.

\section{Numerical integration}

\begin{figure}
\centerline{
\epsfxsize=14.0cm
\epsfbox{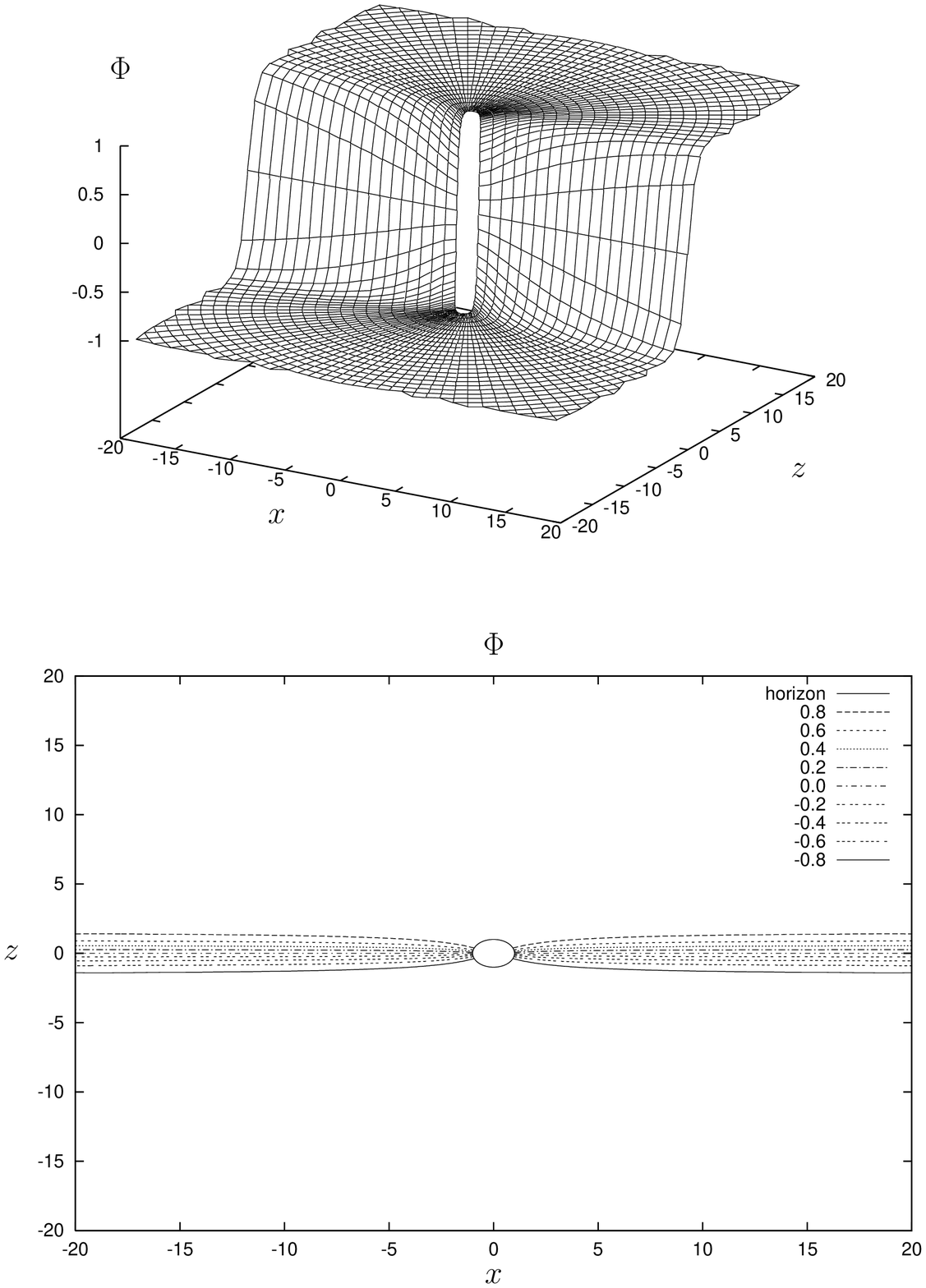}
}
\caption{Numerical solution of Eq.~(\ref{eq:basic}) with the $\phi^4$
potential for $\epsilon=1$.
This solution expresses a domain wall configuration whose thickness is
comparable to the Schwarzschild radius.}
\label{fig:config_phi4_compara}
\end{figure}

\begin{figure}
\centerline{
\epsfxsize=14.0cm
\epsfbox{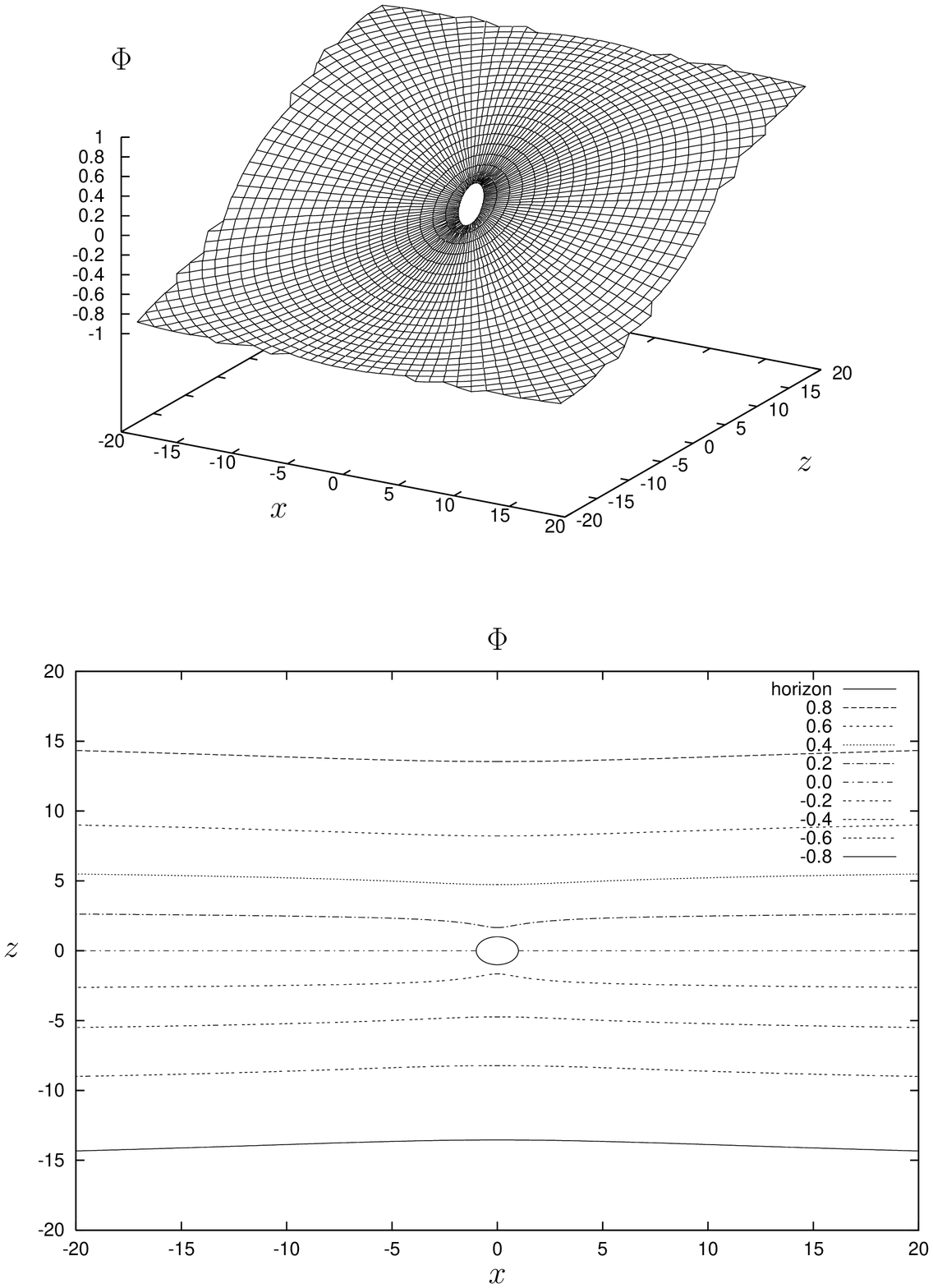}
}
\caption{Numerical solution of Eq.~(\ref{eq:basic}) with the $\phi^4$
potential for $\epsilon=0.1$.
This solution expresses a domain wall configuration whose thickness is
one order of magnitude larger than the Schwarzschild radius.
The black hole is enveloped in the core region of the wall.}
\label{fig:config_phi4_thick}
\end{figure}

\begin{figure}
\centerline{
\epsfxsize=14.0cm
\epsfbox{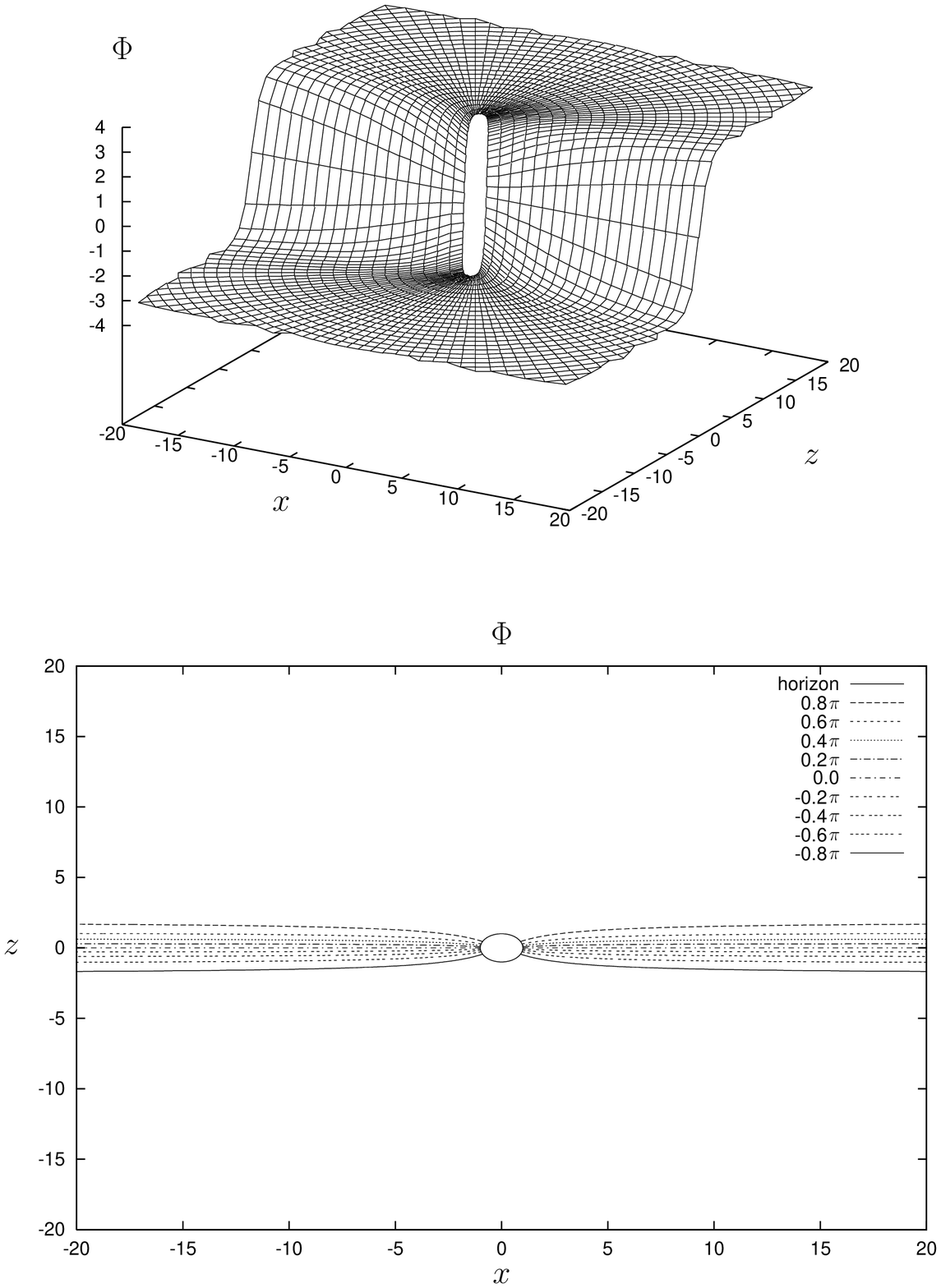}
}
\caption{Numerical solution of Eq.~(\ref{eq:basic}) with the
sine-Gordon potential for $\epsilon=1$.
This solution expresses a domain wall configuration whose thickness is
comparable to the Schwarzschild radius.}
\label{fig:config_sg_compara}
\end{figure}

\begin{figure}
\centerline{
\epsfxsize=14.0cm
\epsfbox{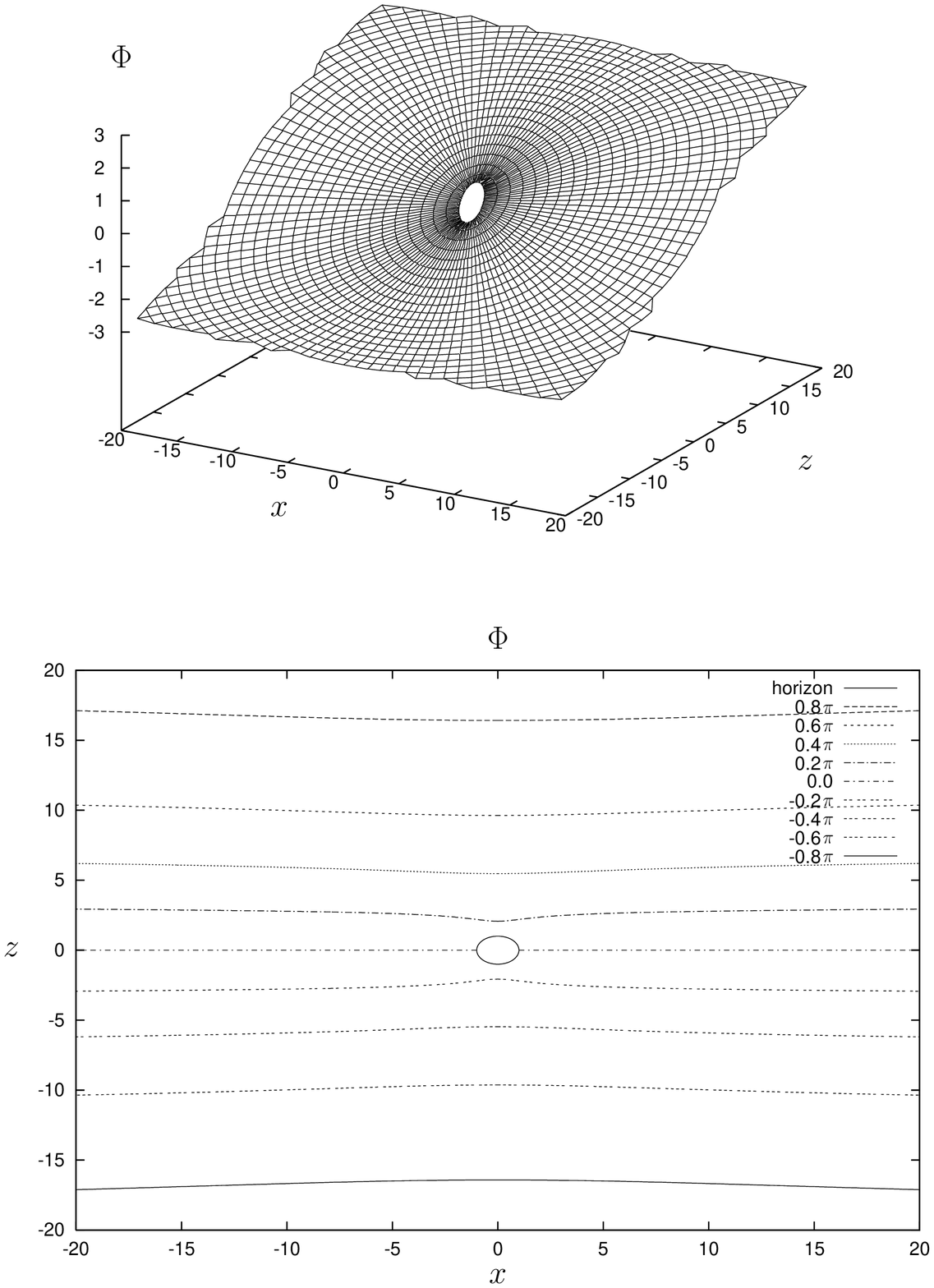}
}
\caption{Numerical solution of Eq.~(\ref{eq:basic}) with the
sine-Gordon potential for $\epsilon=0.1$.
This solution expresses a domain wall configuration whose thickness is
one order of magnitude larger than the Schwarzschild radius.
The black hole is enveloped in the core region of the wall.}
\label{fig:config_sg_thick}
\end{figure}

The scalar field configurations $\Phi(x,z)$ satisfying
(\ref{eq:basic}) and the boundary conditions are shown in
Figs.~\ref{fig:config_phi4_compara}--\ref{fig:config_sg_thick}, where
$x$ and $z$ are the Cartesian coordinates $x=\rho\sin\vartheta$,
$z=\rho\cos\vartheta$.
Here we show the results in typical two cases;
the $\epsilon=1$ case in which the thickness of the kinks
(\ref{eq:BCasymptotic1}), (\ref{eq:BCasymptotic2}) at $\rho_{\rm max}$
is comparable to the Schwarzschild radius
(Fig.~\ref{fig:config_phi4_compara} for $\phi^4$ and
Fig.~\ref{fig:config_sg_compara} for sine-Gordon),
and the $\epsilon=0.1$ case in which the thickness of the kinks
(\ref{eq:BCasymptotic1}), (\ref{eq:BCasymptotic2}) at $\rho_{\rm max}$
is one order of magnitude larger than the Schwarzschild radius
(Fig.~\ref{fig:config_phi4_thick} for $\phi^4$ and
Fig.~\ref{fig:config_sg_thick} for sine-Gordon).
In both the cases, we obtain a domain wall solution as a kink of the
scalar field at the equatorial plane $z=0$.
Particularly in the case $\epsilon=0.1$, the black hole is enveloped
in the core region of the wall.

\begin{figure}
\centerline{
\epsfxsize=14.0cm
\epsfbox{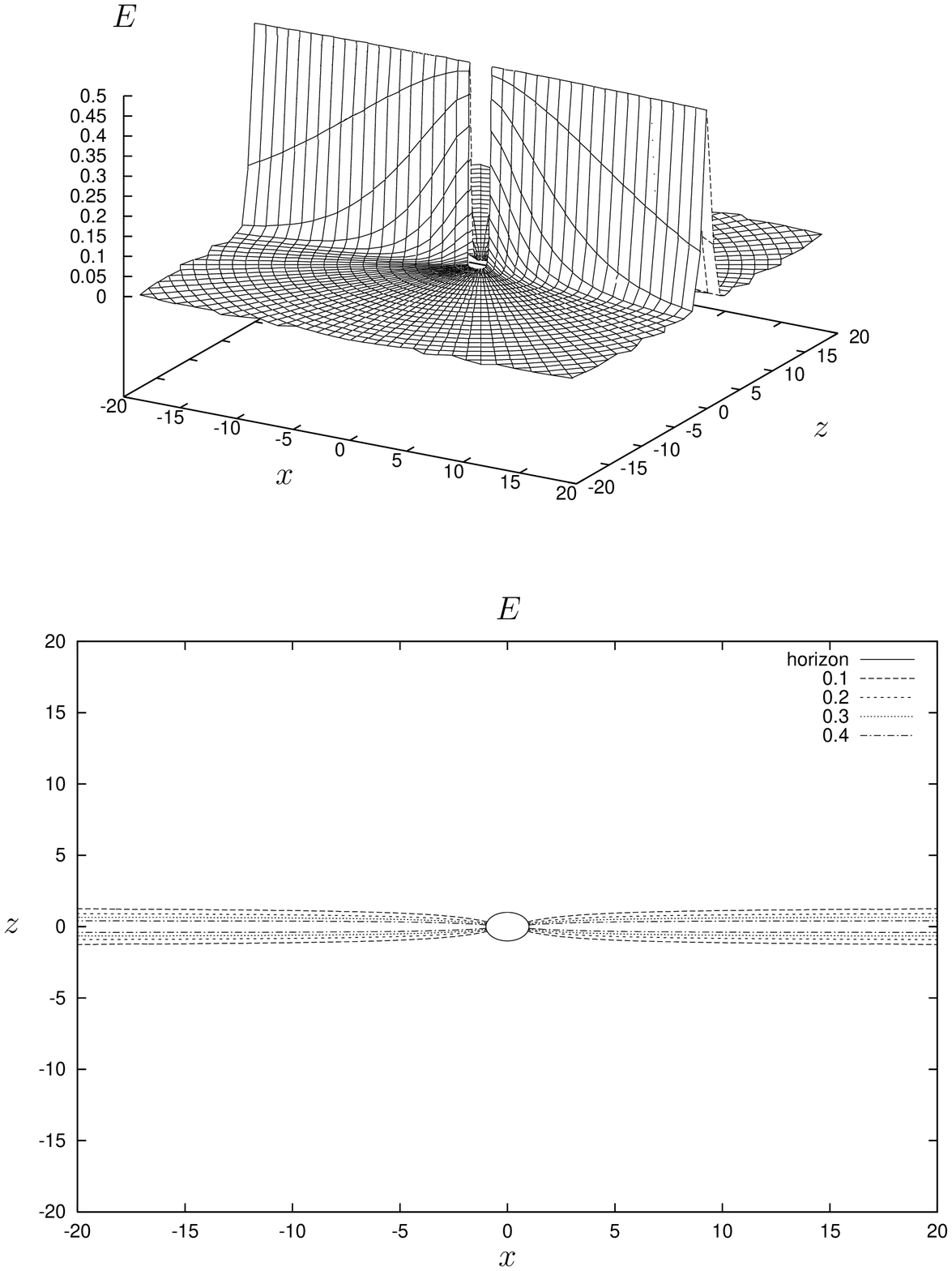}
}
\caption{The energy density $E(x,z)$ of $\phi^4$ scalar field
for $\epsilon=1$.
This is calculated from the numerical solution shown in
Fig.~\ref{fig:config_phi4_compara}}
\label{fig:energy_phi4_compara}
\end{figure}

\begin{figure}
\centerline{
\epsfxsize=14.0cm
\epsfbox{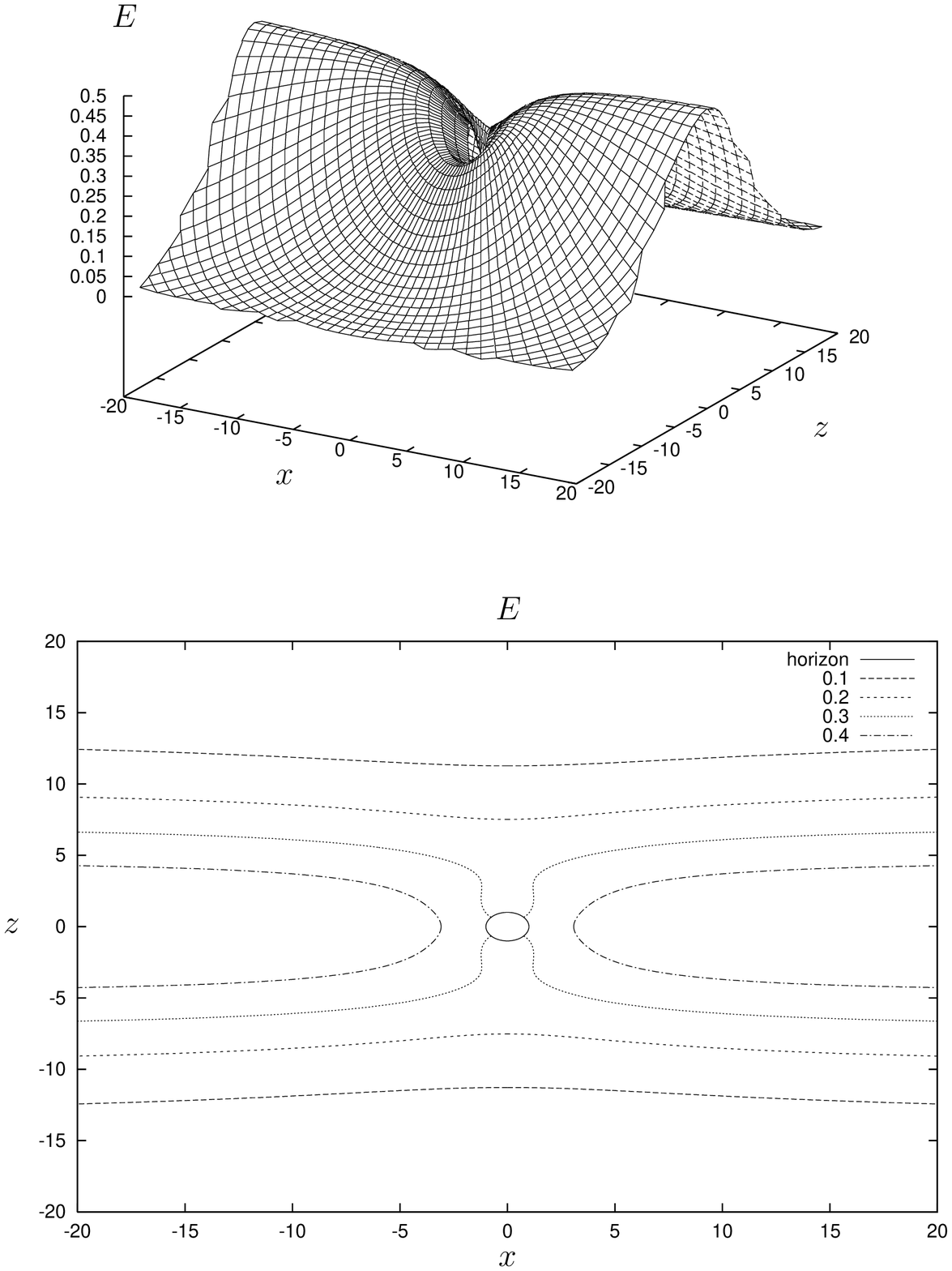}
}
\caption{The energy density $E(x,z)$ of $\phi^4$ scalar field
for $\epsilon=0.1$.
This is calculated from the numerical solution shown in
Fig.~\ref{fig:config_phi4_thick}}
\label{fig:energy_phi4_thick}
\end{figure}

\begin{figure}
\centerline{
\epsfxsize=14.0cm
\epsfbox{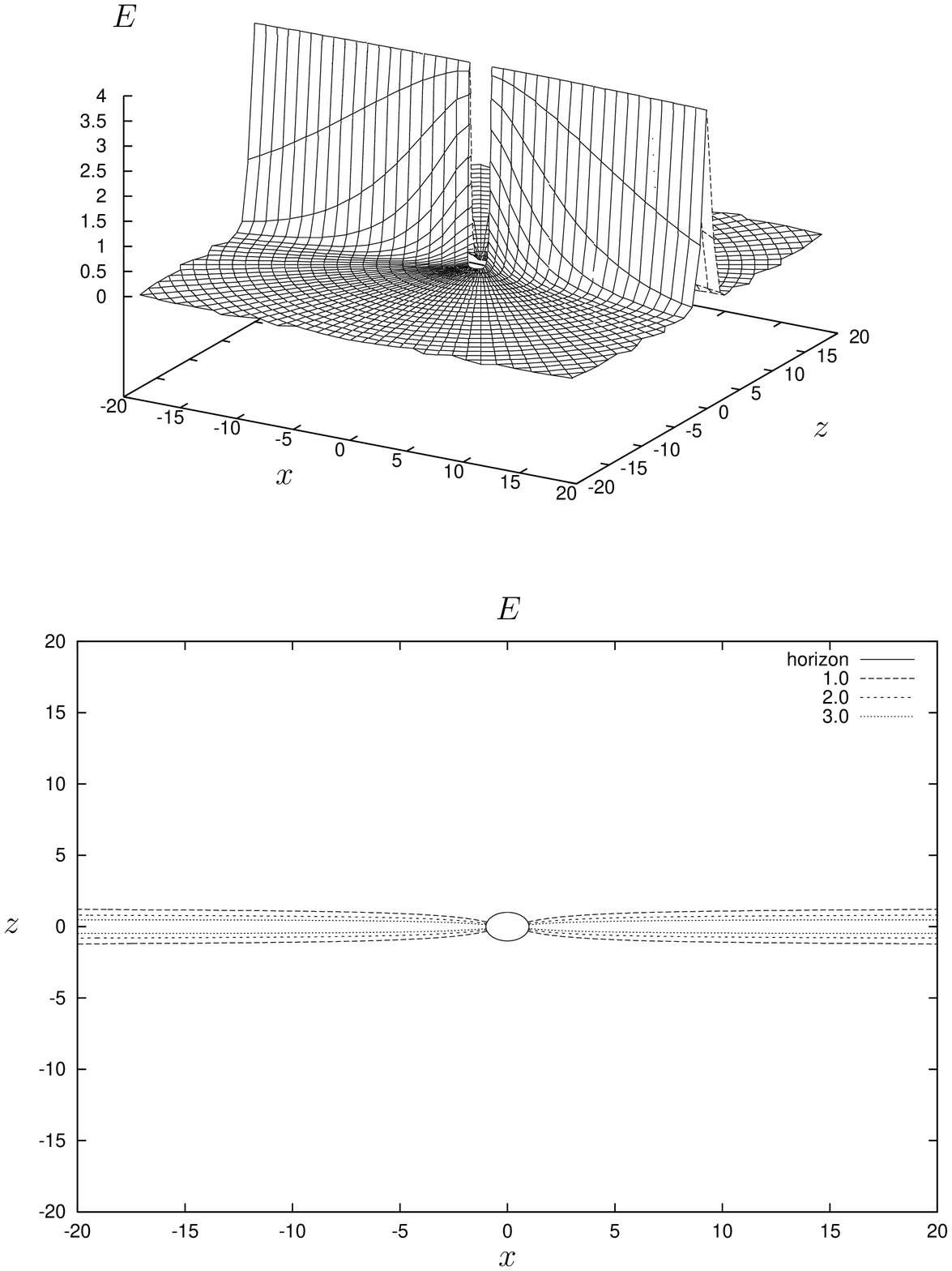}
}
\caption{The energy density $E(x,z)$ of sine-Gordon scalar field for
$\epsilon=1$.
This is calculated from the numerical solution shown in
Fig.~\ref{fig:config_sg_compara}}
\label{fig:energy_sg_compara}
\end{figure}

\begin{figure}
\centerline{
\epsfxsize=14.0cm
\epsfbox{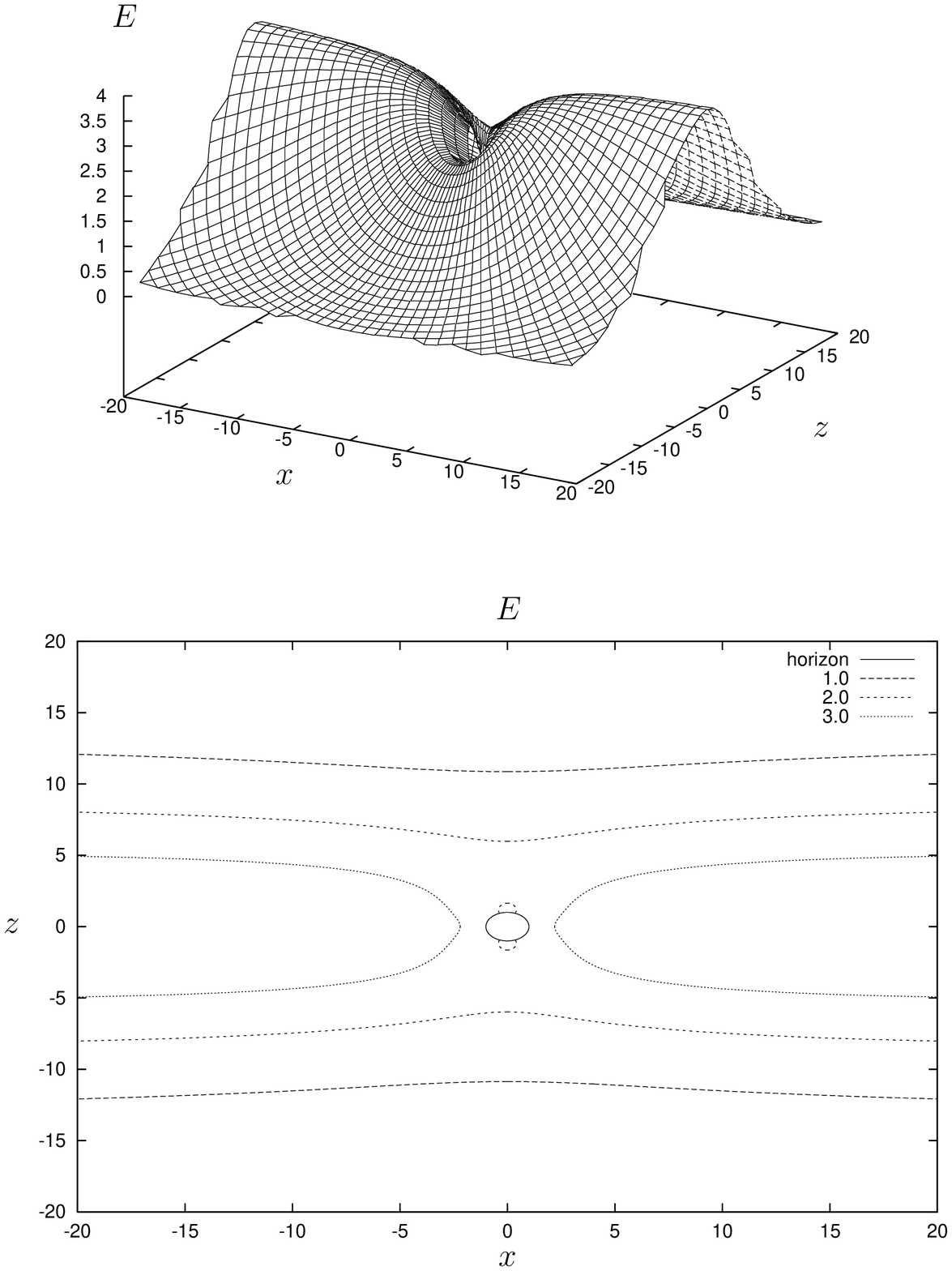}
}
\caption{The energy density $E(x,z)$ of sine-Gordon scalar field for
$\epsilon=0.1$.
This is calculated from the numerical solution shown in
Fig.~\ref{fig:config_sg_thick}}
\label{fig:energy_sg_thick}
\end{figure}

We also show the energy density $E$ of the scalar field given by
\begin{equation}
E \equiv
\frac{T_t{}^t}{\lambda\eta^4}
=
\frac{1}{2\epsilon^2}\left(\frac{\rho}{\rho+1}\right)^{4}
\left[\left(\frac{\partial\Phi}{\partial\rho}\right)^2
+\frac{1}{\rho^2}\left(\frac{\partial\Phi}{\partial\vartheta}\right)^2\right]
+U[\Phi]
\end{equation}
in Figs.~\ref{fig:energy_phi4_compara}--\ref{fig:energy_sg_thick},
corresponding to
Figs.~\ref{fig:config_phi4_compara}--\ref{fig:config_sg_thick},
respectively.
In all the cases, one can see that the configuration actually has a
wall-like structure, namely the energy density is localized around the
equatorial plane with a certain thickness corresponding to $\epsilon$.

We shall comment on the computational domain and the grid spacing
taken in our calculation.
In order that the asymptotic boundary conditions
(\ref{eq:BCasymptotic1}) and (\ref{eq:BCasymptotic2}) make sense,
$\rho_{\rm max}$ must be large enough.
In the above calculation, we take the computational domain which is
fifty times as large as the horizon radius
({\it i.e.} $\rho_{\rm max}=51$), and we carry out the integration on
$500\times90$ grid (the grid spacing in $\rho$- and $\vartheta$-
direction is $0.1\times$horizon-radius and $1^{\circ}$, respectively).
Then we clip the region $\{|x|,|z|<20\}$, where the above results are
insensitive to the value of $\rho_{\rm max}$ and the number of grid
points.
In fact, the results do not change when we extend the computational
domain to $\rho_{\rm max}=101$ with keeping the grid spacing, and the
results differ from ones on the finer ($1000\times180$) grid with
keeping $\rho_{\rm max}=51$ at most 1\%.
We also comment that the reliability of the numerical code is checked
in the flat space case and it reprodeces the exact solutions
(\ref{eq:flat4}), (\ref{eq:flatSG}) with accuracy of $10^{-2}$.

\section{Summary and Discussions}

In order to answer the question whether or not scalar fields can actually 
form a topological defect in the vicinity of a black hole, 
we have numerically solved the equation of motion for real scalar fields 
with $\phi^4$ and sine-Gordon potentials, 
which have a discrete set of degenerate minima, 
in the Schwarzschild black hole background.  
In both $\phi^4$ and sine-Gordon potential cases, 
we showed that there exist the static axi-symmetric field configurations 
which represent thick domain walls intersecting the black hole.
In particular, we studied the specific case that the wall's core is located 
at the equatorial plane of the Schwarzschild spacetime.  
We introduced the parameter $\epsilon$ which characterizes 
the domain wall thickness compared to the black hole horizon radius; 
the smaller than unity $\epsilon$ is, the larger the wall width is. 
We showed the domain wall solutions and their energy densities 
for $\epsilon =1$ and $\epsilon =0.1$ cases. 
In summary, we can say that a black hole is not an obstacle 
for scalar fields to form a domain wall configuration intersecting 
the black hole. 

One might wonder about our present results; 
one naively expects that the scalar field could have no static 
distribution around a black hole and inevitably fall into the horizon
as usual objects do.
However, a domain wall is a relativistic object with a large negative
pressure (or large tension) whose magnitude is comparable to that of
energy density.
Furthermore in our study we examined the domain walls which are
extended infinitely in space. 
Then we can understand that the domain wall is suspended from the
asymptotic region and supported against falling into the black hole
by its tension, so that the static configuration is realized. 

In our analysis, we assumed that the gravitational effect of the domain wall 
is negligible compared to that of the Schwarzschild black hole. 
We shall comment on the validity of this assumption. 
The energy density of a domain wall is given by 
\begin{equation}    
GT_t{}^t \sim G\lambda\eta^4
\sim \frac{1}{w^2}\left(\frac{\eta}{m_{\rm Pl}}\right)^2,
\end{equation}
where $m_{\rm Pl}$ is the Planck mass and $G$ is the Newtonian
constant.
On the other hand, the curvature strength, or gravitational tidal force, 
of Schwarzschild spacetime is estimated as 
\begin{equation}
\frac{GM}{R^3} \sim \epsilon \frac{w}{R^3} 
\end{equation}
at areal radius $R$. 
Then a ratio of the gravity which will be produced by the domain wall
to the gravity of the background black hole is given by
\begin{equation}
\omega \equiv \frac{GT_t{}^t}{GM/R^3}
\sim
\frac{1}{\epsilon}\left(\frac{R}{w}\right)^3
\left(\frac{\eta}{m_{\rm Pl}}\right)^2.
\label{eq:gravitating_param}
\end{equation}
When $\omega$ becomes much smaller than unity, the gravity of the
domain wall is negligible compared to that of the black hole, and our
test wall assumption becomes valid.
From Eq. (\ref{eq:gravitating_param}) we have 
$\omega \sim \epsilon^2(\eta/m_{\rm Pl})^2$
near the horizon ($R\sim 2GM$).
Therefore, when we consider domain walls with the symmetry breaking
scale being much lower than the Planck scale
({\it i.e.} $\eta\ll m_{\rm Pl}$), we have $\omega\ll 1$ and
consequently our result gives a good description of shapes of
gravitating thick domain walls near the black hole.
We can also see from Eq.(\ref{eq:gravitating_param}) that, in the
thick wall ($w > 2GM$) case, if $\eta\ll m_{\rm Pl}$, the test wall
assumption is still valid even at $R\sim w$.

In the asymptotic region ($R\gg w$), one may expect that 
the gravity of the domain wall is no longer negligible and changes 
the asymptotic geometry drastically. 
Bonjour, Charmousis and Gregory have recently investigated 
the spacetime of a thick gravitating domain wall with 
local planar symmetry and reflection symmetry around the wall's 
core~\cite{Bonjour:1999kz}. They showed that the domain wall spacetime 
becomes spatially compact and has a cosmological horizon as de Sitter 
spacetime does. 
This suggests that, when a black hole exists and wall's gravity 
is taken into account in the region far from the black hole, 
the whole spacetime has a cosmological horizon 
and an axi-symmetric domain wall intersects both the black hole and 
the cosmological horizons 
as an equatorial plane in a Schwarzschild-de Sitter spacetime. 
This motivates us to study further the interaction between 
thick domain walls and black holes 
in, for example, Schwarzschild-de Sitter background.

The domain wall solutions obtained here are thought to represent 
a possible final configuration of 
a gravitational capturing of a domain wall by a black hole. 
At present, it is far reaching for us to investigate a fully dynamical 
process such as the scattering and capture of thick domain walls 
by black holes. 
However, to get some insights into the problem, 
it is worth investigating the existence of the static axi-symmetric 
solutions which represent thick domain walls located away from a black hole.

A cosmic string is also an extended relativistic object with large tension 
and thought to play more important role in cosmology than a domain wall does. 
Study of a gravitationally interacting system of thick cosmic strings 
and black holes is an interesting problem as a generalization of 
the present analysis. 

\acknowledgments

We would like to thank Profs. Hideki~Ishihara, Hideo~Kodama and
Takashi~Nakamura for many useful suggestions and comments.
We also thank Susumu~Higaki for stimulating discussions.
D.I. was supported by JSPS Research Fellowships for Young Scientists,
and this work was supported in part by the Grant-in-Aid for Scientific
Research Fund (No. 4318).
A.I. was supported by JSPS Research Fellowships for Young Scientists 
and this work was partially supported by Handai Yukawa Shogakukai (A.I.). 

\appendix

\section{The Boundary Condition at the Event Horizon}
The tangent of a freely falling observer parameterized by its proper
time $\tau$ is $u^{\mu}=({\rm d}t/{\rm d}\tau,{\rm d}r/{\rm d}\tau,0,0)$, and we have
\begin{equation}
-1=g_{\mu\nu}u^{\mu}u^{\nu}=
-\left(\frac{2r-M}{2r+M}\right)^2\left(\frac{{\rm d}t}{{\rm d}\tau}\right)^2
+\left( 1+{M \over 2r} \right)^4 
\left(\frac{{\rm d}r}{{\rm d}\tau}\right)^2.
\end{equation}
The quantity
\begin{equation}
\alpha\equiv -g_{\mu\nu}\xi^{\mu}u^{\nu}=
\left(\frac{2r-M}{2r+M}\right)^2\left(\frac{{\rm d}t}{{\rm d}\tau}\right)
\end{equation}
is a constant of the motion, where $\xi^{\mu}=(1,0,0,0)$ is the static
Killing field.
The energy density observed by this observer is
\begin{equation}
T_{\mu\nu}u^{\mu}u^{\nu} =
-(u\cdot\nabla\phi)^2
-\left(\frac{1}{2}\nabla\phi\cdot\nabla\phi+V[\phi]\right).
\label{eq:Tuu}
\end{equation}
Since we consider the static configuration and the spatial part of the 
metric is non-singular in the isotropic coordinates, the second term
of Eq.~(\ref{eq:Tuu}) is always finite.
We have
\begin{equation}
u\cdot\nabla\phi= - \left( 1+ {M \over 2r} \right)^{-2}
\left[\alpha^2\left(\frac{2r+M}{2r-M}\right)^2-1\right]^{1/2}
\frac{\partial\phi}{\partial r}
\end{equation}
for the static axi-symmetric configuration $\phi(r,\vartheta)$.
Thus the requirement that Eq.~(\ref{eq:Tuu}) is finite at the horizon
is reduced to $\partial\phi/\partial r=0$ at $r=M/2$, or equivalently
Eq.~(\ref{eq:BChorizon}).

\begin{references}
\bibitem{VilenkinBook} A.~Vilenkin and E.P.S.~Shellard, {\it Cosmic
Strings and other Topological Defects} (Cambridge University Press,
New York, 1994).

\bibitem{Hill:1989vm}
C.~T.~Hill, D.~N.~Schramm and J.~N.~Fry,
Comments Nucl.\ Part.\ Phys.\  {\bf 19}, 25 (1989).

\bibitem{Vilenkin:1983hy}
A.~Vilenkin,
Phys.\ Lett.\  {\bf B133}, 177 (1983).

\bibitem{Ipser:1984db}
J.~Ipser and P.~Sikivie,
Phys.\ Rev.\  {\bf D30}, 712 (1984).

\bibitem{DeVilliers:1998nk}
J.~De Villiers and V.~Frolov,
Int.\ J.\ Mod.\ Phys.\  {\bf D7}, 957 (1998)
[gr-qc/9711045].

\bibitem{DeVilliers:1998xz}
J.~De Villiers and V.~Frolov,
Phys.\ Rev.\  {\bf D58}, 105018 (1998)
[gr-qc/9804087].

\bibitem{Page:1998ya}
D.~N.~Page,
Phys.\ Rev.\  {\bf D58}, 105026 (1998)
[gr-qc/9804088].

\bibitem{DeVilliers:1999nm}
J.~De Villiers and V.~Frolov,
Class.\ Quant.\ Grav.\  {\bf 16}, 2403 (1999)
[gr-qc/9812016].

\bibitem{Page:1999fd}
D.~N.~Page,
Phys.\ Rev.\  {\bf D60}, 023510 (1999)
[gr-qc/9902038].

\bibitem{Christensen:1998hg}
M.~Christensen, V.~P.~Frolov and A.~L.~Larsen,
Phys.\ Rev.\  {\bf D58}, 085008 (1998)
[hep-th/9803158].

\bibitem{Frolov:1999td}
V.~P.~Frolov, A.~L.~Larsen and M.~Christensen,
Phys.\ Rev.\  {\bf D59}, 125008 (1999)
[hep-th/9811148].

\bibitem{Cline:1998ft}
For an overview, {\it Proceedings, 3rd International Symposium, and
Workshop on Primordial  Black Holes and Hawking Radiation, 1998},
edited by D.~B.~Cline,
Phys.\ Rept.\  {\bf 307}, 1 (1998).

\bibitem{Carr:YKIS99}
B.~Carr and C.~Goymer,
Prog.\ Theor.\ Phys.\ Suppl.\  {\bf 136}, 321 (1999)

\bibitem{Yokoyama:YKIS99}
J.~Yokoyama,
Prog.\ Theor.\ Phys.\ Suppl.\  {\bf 136}, 338 (1999)

\bibitem{Battye:1999eq}
R.~A.~Battye, M.~Bucher and D.~Spergel,
astro-ph/9908047.

\bibitem{MYII2}
Y.~Morisawa,~R.~Yamazaki,~D.~Ida, A.~Ishibashi, and K.~Nakao, in preperation.

\bibitem{Bonjour:1999kz}
F.~Bonjour, C.~Charmousis and R.~Gregory,
Class.\ Quant.\ Grav.\  {\bf 16}, 2427 (1999)
[gr-qc/9902081].
\end{references}
\end{document}